\documentclass[pdflatex,sn-nature]{sn-jnl}
\usepackage{graphicx}%
\usepackage{multirow}%
\usepackage{amsmath,amssymb,amsfonts}%
\usepackage{amsthm}%
\usepackage{mathrsfs}%
\usepackage[title]{appendix}%
\usepackage{xcolor}%
\usepackage{textcomp}%
\usepackage{manyfoot}%
\usepackage{booktabs}%
\usepackage{algorithm}%
\usepackage{algorithmicx}%
\usepackage{algpseudocode}%
\usepackage{listings}%

\UseRawInputEncoding

\begin{document}

\title[Massive black holes in the first billion years]{Massive black holes in the first billion years with {\it JWST}}

\author{\fnm{Hannah} \sur{\"Ubler}}

\affil{\orgdiv{Max-Planck-Institut f\"ur extraterrestrische Physik}, \orgaddress{\street{Gie{\ss}enbachstra{\ss}e 1}, \city{Garching}, \postcode{85748}, \country{Germany}};  \texttt{hannah@mpe.mpg.de}}

\abstract{The {\it James Webb Space Telescope} has unlocked uncharted territory for the study of early black holes, enabling the detection of fainter and smaller black holes at larger cosmological distances than previously possible. Compared to nearby massive black holes, they are characterised by several unexpected attributes, such as their multi-wavelength properties or their relation to their host galaxies. These observational findings challenge our understanding of the past evolution of present-day supermassive black holes, and provide novel anchor points to constrain theoretical models of black hole formation and growth. This review summarises our current understanding of the observed properties of newly discovered, active black holes in the early Universe, discusses the status of mass measurements and the relation to their hosts, as well as open questions and future prospects.
}

\maketitle

Supermassive black holes (SMBHs) are ubiquitous in the centres of massive galaxies in the local Universe \citep{Magorrian98, Kormendy13}, such as the Milky Way \citep{Eckart96, Ghez98}, yet their birth and growth mechanisms remain poorly understood \citep{Inayoshi20}. 
Starting at a million solar masses ($M_\odot$) in the `supermassive' category, the most massive black hole with a dynamical mass measurement in the local Universe weighs 40 billion solar masses \citep{Mehrgan19}. 
The key pathways to forming such massive black holes have been identified long ago \citep{Rees84}. The consensus is that SMBHs have started out with initial (`seed') masses that are typically orders of magnitude lower \citep{Inayoshi20, Volonteri21}. Candidate seeds range from stellar remnants ($\sim10~M_\odot$; including potentially more massive remnants of the first generation of stars, $\sim10^{2-3}~M_\odot$; e.g.~\citealp{Heger03}), runaway collisions in dense star clusters ($\sim10^4 M_\odot$; e.g.~\citealp{Begelman78}), `direct collapse black holes' (DCBHs) formed out of supermassive stars in atomic cooling haloes ($\sim10^{4-6}~M_\odot$; e.g.~\citealp{Haehnelt93}), to primordial black holes formed out of density fluctuations at the beginning of the Universe with a wide range of possible masses ($\lesssim10^5~M_\odot$; \citealp[e.g.][]{Carr20}). 
Despite substantial theoretical efforts, current simulations still have limited ability to constrain the most likely seeding process \citep{Inayoshi20, Volonteri21}.

The question of black hole seeding has become more pressing in recent decades through the discovery of hundreds of luminous quasars at redshift $z>6$, when the Universe was less than one billion years old \citep[see][for a recent review]{Fan23}.
How did these black holes manage to grow so massive so quickly? Does their existence demonstrate already that SMBHs are born heavy? Not quite; the picture is more complicated since also the growth modes of black holes are not well constrained. For instance, numerous observations show that accretion above the theoretical `Eddington limit' is not uncommon \citep[e.g.][]{Fan23}. Besides, their high luminosities identify quasars as a {\it rare} class of objects, the brightest black holes. 

Accreting black holes, so called `active galactic nuclei' (AGN), with lower luminosities make up the bulk of the active black hole population. They can be identified with the help of rest-frame optical lines: from permitted broad-line emission (the broad-line region, BLR) associated with the fast, FWHM~$\gtrsim1000$~km~s$^{-1}$ motions of gas around massive black holes (Type 1 AGN); from high-velocity, blue-shifted nuclear emission in permitted and forbidden lines, indicative of powerful AGN-driven outflows; or from narrow (FWHM~$\lesssim400$~km~s$^{-1}$) emission line ratio diagnostics indicating dominant ionisation by an AGN (called Type 2 AGN in the absence of a BLR).
AGN population properties and their impact on galaxies have been characterised up to $z\sim3$ through ground-based surveys targeting their rest-frame optical emission \citep[e.g.][]{FS19, Kakkad20}. 
However, because these wavelengths are redshifted out of Earth's atmospheric windows for galaxies beyond $z\sim3.5$, only individual AGN had been detected at earlier cosmic times through different spectroscopic or photometric signatures.

Thus, the advent of a new observing machine in the sky which could detect the lower-luminosity, more typical AGN population at early cosmic times had been eagerly awaited: the {\it James Webb Space Telescope (JWST)} \citep{Gardner06}. And it delivers. Thanks to sensitive near-to-mid-infrared spectroscopic instruments \citep{Jakobsen22, Argyriou23}, {\it JWST} routinely observes galaxies and active black holes at cosmic dawn.
The observed properties of these fainter black holes, and what they tell us about early black hole growth, are the subject of this review.

\section*{The new discovery space with JWST}\label{}

Right from the beginning of science operations, {\it JWST} revealed hitherto hidden black holes in the early Universe \citep{Kocevski23}. The game changer was the ability of spectroscopic instruments aboard {\it JWST} to observe rest-frame optical emission above $z\sim3.5$. 
Taking the nominal wavelength range of the NIRSpec instrument as an example, we can now observe H$\alpha$, the brightest hydrogen line, up to redshifts of $z\sim7$, and H$\beta$ up to $z\sim9.9$ (less than 500~Myr after the Big Bang). Another crucial aspect is {\it JWST's} sensitivity, unaffected by atmospheric emission, able to reveal objects at a fixed redshift much fainter than previously possible. Taken together, these features define the new parameter space opened up by {\it JWST}, not just for the study of black holes: fainter and farther.

\begin{figure}[h]
\centering
\includegraphics[width=0.95\textwidth]{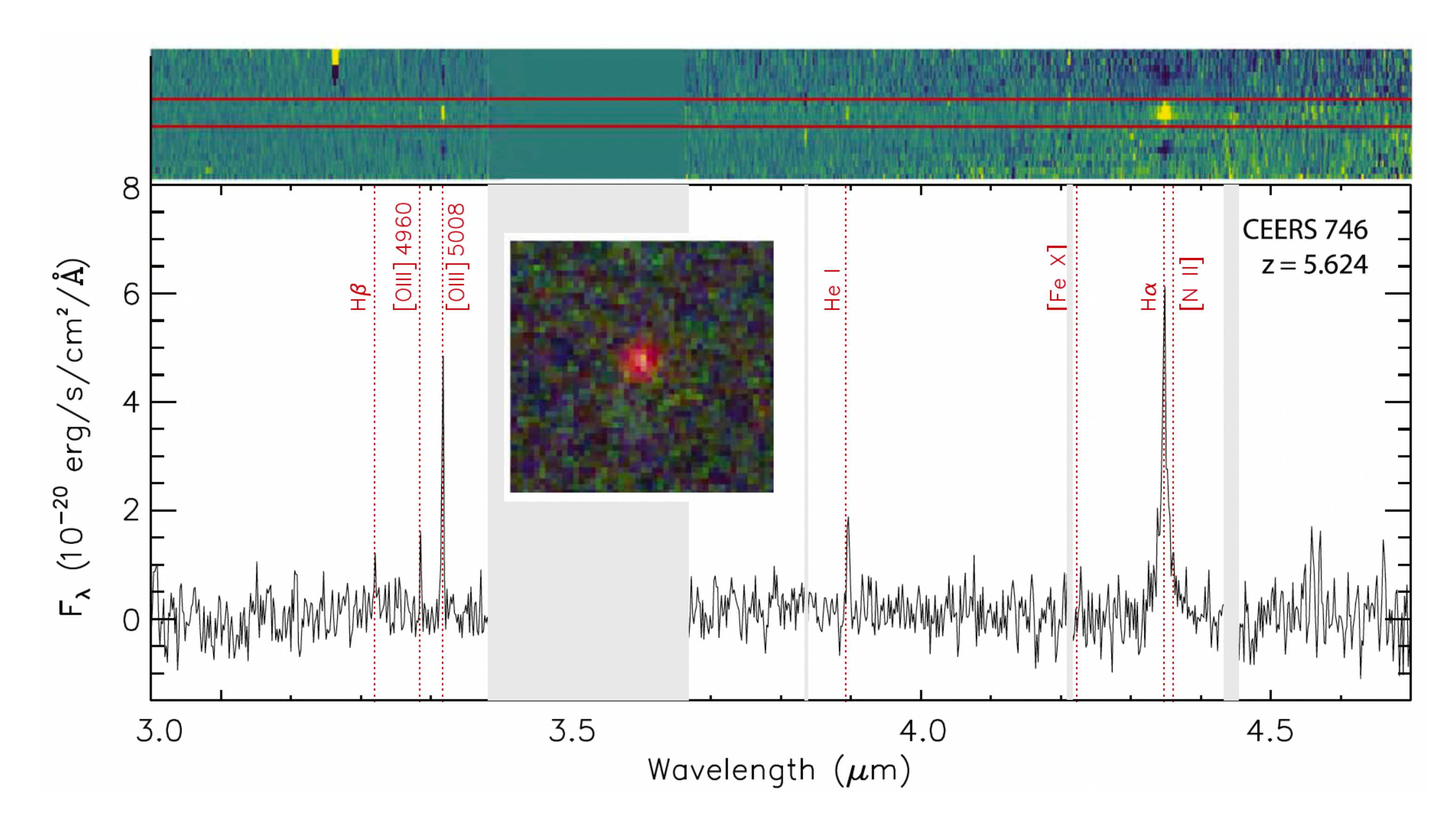}
\caption{One of the first broad-line AGN detected with {\it JWST}. The two-dimensional NIRSpec $R\sim1000$ spectrum is shown at the top; the one-dimensional spectrum (bottom) is extracted from the region highlighted in red. A $2^{\prime\prime}\times2^{\prime\prime}$ NIRCam rgb image (F150W, F277W, F444W) is shown in the inset  (12~kpc$\times$12~kpc at $z\sim5.5$), highlighting the small size of the source and the prominent flux contribution in the reddest filter.
Narrow [O{\sc{iii}}]$\lambda\lambda4960,5008$ emission is visible whereas H$\alpha$ features also a broad component, indicative of the BLR of an active black hole. [N{\sc{ii}}]$\lambda\lambda6550,6585$ is not significantly detected. 
Figure adapted under a \href{https://creativecommons.org/licenses/by/4.0/}{Creative Commons Attribution 4.0 license} from ref.~\citep{Kocevski23}.}\label{f:agn_example}
\end{figure}

The first `new' AGN were detected at $z\gtrsim5$ based on NIRSpec spectroscopy or NIRCam grism data, revealing their BLR emission in hydrogen lines underneath a narrow component, whereas bright forbidden lines such as [O{\sc{iii}}]$\lambda5008$ showed only narrow emission \citep[e.g.][]{Kocevski23, Uebler23, Larson23, Kokorev23, Greene24, Matthee24, Maiolino24}. 
Some of these sources had already been identified as AGN candidates based on the detection of high-ionisation rest-frame UV lines or spectral energy distribution (SED) fitting \citep[e.g.][]{Grazian20, Barchiesi22, Onoue23}. Indeed, the confirmation and detection of many AGN was made possible through programs combining imaging with spectroscopic follow-up \citep[e.g.][]{Bezanson24, Finkelstein25, Eisenstein26}.
Morphologically, the sources were mostly compact, particularly in the rest-frame optical where the BLR emission dominates. 
Spectroscopy showed that the narrow emission components of these AGN were compatible with low metallicity ($Z\sim0.2-0.4~ Z_\odot$). 
The [N{\sc{ii}}]$\lambda\lambda6550,6585$ doublet was weak or not detected, while some high-ionisation lines like He{\sc{ii}}$\lambda4868$ or tentative detections of coronal lines were reported \citep{Kocevski23, Uebler23}.
We show one of the first two {\it JWST} Type 1 AGN discovered at $z\gtrsim5$ in Fig.~\ref{f:agn_example}.
Within months, numerous additional Type~1 AGN were discovered \citep[e.g.][]{Uebler23, Larson23, Kokorev23, Greene24, Matthee24, Maiolino24}.

\paragraph{The challenging identification of Type 2 AGN}
It quickly became clear that Type 2 AGN would be harder to find \citep{Kocevski23, Uebler23}. Classical emission line ratio diagnostics are effective in separating nearby AGN from galaxies dominated by ionisation from stars \citep{Baldwin81, Veilleux87}. However, the physical conditions in distant galaxies hamper this distinction: in the BPT diagram \citep{Baldwin81}, star-forming galaxies move towards higher [N{\sc{ii}}]/H$\alpha$ and [O{\sc{iii}}]/H$\beta$ due to harder ionising radiation fields and larger electron densities, while low-metallicity AGN move towards lower [N{\sc{ii}}]/H$\alpha$, and the two populations start to overlap \citep{Kewley13}.  
This effect is least severe for [O{\sc{iii}}]/H$\beta$ versus [O{\sc{i}}]$\lambda6300$/H$\alpha$ (\citealp{Veilleux87}; see \citealp{Juodzbalis25}). 
In response to this, more conservative cuts were introduced for classical diagrams, designed to select robust, high-metallicity AGN \citep{Scholtz25}. New rest-frame optical diagnostics utilising [O{\sc{iii}}]$\lambda4363$/H$\gamma$ were proposed which identify robust AGN based on the higher average energy of ionising photons from an AGN compared to stars \citep{Mazzolari24}. These diagrams achieve a high purity in AGN selection, but lack completeness especially for systems with low metallicity. Such AGN can be identified with rest-frame UV diagnostics, particularly those involving He{\sc{ii}}$\lambda1640$ which is more easily produced by the harder ionising radiation field of AGN (\citealp{Feltre16, Hirschmann23}; also \citealp{Nakajima22}). Currently, a limitation in systematically applying these diagnostics is a scarcity of high-spectral resolution rest-frame UV coverage of distant galaxies \citep{Scholtz25}. 
In the rest-frame optical instead, He{\sc{ii}}$\lambda4686$ as used in the metallicity-independent diagnostic proposed by \cite{Shirazi12} may help to correctly identify AGN \citep{Uebler23}, but is rarely detected in distant galaxies.
Notwithstanding, systematic efforts in identifying Type 2 AGN mainly in the GOODS-S and CEERS fields have produced about 100 sources \citep{Scholtz25, Mazzolari25}. These works employ a combination of diagnostic diagrams and consider a galaxy as AGN when robustly classified as such in at least one diagnostic.\\  

So far, several hundred AGN have been detected with {\it JWST}. More than 300 have a black hole mass reported, of which more than 200 are at $z\gtrsim4$. The most distant AGN with a black hole mass estimate to date is found at $z=10.6$ \citep{Maiolino23a}.
Among the new AGN, a subclass emerged dubbed `Little Red Dots' \cite{Matthee24}. These sources are red, compact (unresolved with the {\it JWST} point-spread function at rest-frame optical wavelengths), broad-line AGN with a negative spectral slope blue-wards and a positive slope red-wards of the Balmer limit at $\lambda=3646$~\AA\ (the `v-shape'), which account for 30-40 per cent of the newly detected BLR AGN \citep[][]{Greene24, Hainline25, Hviding25}. 
In this review, we discuss the new AGN population revealed by {\it JWST}, referring to LRDs where appropriate. For a specific discussion of LRDs we refer the reader to \cite{Inayoshi26}.

\section*{Unexpected properties of distant AGN}\label{}

With the first {\it JWST} discoveries of distant AGN came also the first surprises. In the following, the main unexpected commonalities of {\it JWST} AGN are discussed.

{\bf X-ray and radio weakness. } Although many {\it JWST} AGN are located in deep {\it Chandra} fields, they are undetected in $X-$rays \citep{Ananna24, Yue24, Maiolino25a}, with few individual exceptions (e.g.~\citealp{Maiolino25a, Kocevski25, Napolitano25, Hviding26, Bogdan24}, but see \citealp{Zou26}). This could be due, for instance, to absorption \citep[e.g.][]{Maiolino25a} or super-Eddington accretion \citep[e.g.][]{Pacucci24, Lambrides24}.
This has been a surprising result, since $X-$rays are considered a tell-tale signature of accreting black holes. In fact, the black hole accretion rate density in the early Universe has been largely constrained through $X-$ray observations and shows a shape remarkably similar to that of the cosmic star-formation rate density \citep{Marconi04, Madau14}. Therefore, no matter the cause of the $X-$ray weakness, there is an immediate implication: one pillar of the `co-evolution paradigm' between black holes and galaxies begins to totter. 
{\it JWST} AGN appear also weak at radio frequencies \citep[e.g.][]{Gloudemans25, Mazzolari26a}. However, current upper limits are still compatible with expectations for radio-quiet AGN.

{\bf Lack of strong outflows. } 
Another curiosity is a lack of strong outflow emission in the rest-frame optical, typically associated with AGN at lower redshifts and with quasars. Only few of the high$-z$ lower-luminosity AGN display outflow signatures in emission \citep[e.g.][]{Uebler23, Maiolino23a, Korber26}. This could be due to their low metallicity \cite{Maiolino25a}, but a dedicated investigation is still pending.

{\bf Lack of variability. } 
{\it JWST} AGN lack variability. This is unexpected because variability is a key feature of lower-redshift AGN and quasars \citep[e.g.][]{Ulrich97}. So far, only few cases of (tentative) continuum or spectral variability have been reported \citep{Ji25, Naidu25, ZhangZ25, DEugenio25b}, while in most multi-epoch observations no variability is detected \citep[e.g.][]{Uebler24b, Kokubo25, Stone25, Tee25, Liu26}.
This lack could be due to super-Eddington accretion \citep{Secunda26}, or could point towards longer variability timescales related to higher black hole masses \citep[see e.g.][]{MacLeod10}.
Of course, we need to keep in mind that the timescales probed by {\it JWST} are still limited for unlensed sources: maximally about 5-10 rest-frame months for sources at $z\sim4-8$ considering the current mission length of about four years.

{\bf LRDs. } 
The properties listed above apply broadly to the new AGN population, including LRDs. Some features, however, are unique to LRDs, which we briefly discuss here \citep[see][for further details]{Inayoshi26}. 
Balmer absorption was first detected in the broad H$\alpha$ emission of two sources at $z\sim5-5.5$, indicative of high neutral hydrogen densities ($N_H\gtrsim10^{19}$~cm$^{-2}$, $n_H\gtrsim10^{9}$~cm$^{-3}$) within or very close to the BLR \cite{Matthee24}. This phenomenon, fairly rare at $z<2$, turned out to be common among LRDs \citep[$> 20 \%$; e.g.][]{Lin24, Kocevski25}.
The high occurrence of hydrogen (and helium) absorption immediately suggests that the covering factor of the absorbing medium around the black hole must be high, distinguishing it from the typical geometry of lower$-z$ BLRs. 
It was soon postulated that the peculiar spectral shape of LRDs could arise as a consequence of absorption in the Balmer series by high-density gas clouds along the line of sight \cite{Inayoshi25}, and the application of this concept to data was first demonstrated for a $z\sim7$ LRD (\citealp{Ji25}; see also \citealp{Naidu25, deGraaff25a, Taylor25b}). 
Subsequently, it was empirically argued that the majority of LRDs can be modelled as AGN within a thermalised dense gas envelope \cite{Kido25, deGraaff25}.
These concepts share similarities with the original idea of a
`quasistar' \cite{Begelman06}: a black hole collapsed out of a dense gas cloud or supermassive star, initially small but rapidly growing by (super-Eddington) accretion from the surrounding envelope \citep[see also][]{Begelman26}. 
While a fully satisfactory description {of LRD properties} within a cosmological framework is still pending, there is broad agreement within the community that the existence of dense gas near the black hole is required to explain the spectral properties of LRDs.

It is still not clear whether LRDs represent a distinct class of Type I AGN or a specific phase of black hole evolution. Recently, it was argued that LRDs are no different from the population of `normal' {\it JWST} AGN, but are simply observed in a different evolutionary stage and/or under a different angle \citep{Brazzini26, Matthee26, Madau26}, reminiscent of past discussions surrounding Type~1 and Type~2 AGN.
To make progress, it will be important to characterise the demographics of all AGN across cosmic time.

\section*{How to weigh the first black holes}\label{}

Another unexpected property of {\it JWST} AGN deserves attention: their black hole masses. At lower redshifts accurate black hole masses can be obtained through reverberation mapping of light echos from active black holes or through stellar or gas dynamics, most notably for the Milky Way SMBH through orbits of individual stars \citep[][and references therein]{Genzel24}. 
At $z>2$ instead we largely rely on estimates based on local scaling relations between BLR width, line or continuum luminosity, and black hole mass \citep[e.g.][]{Greene05, Vestergaard06}, called the `single-epoch' method (SE). This method is limited to mass estimates for active black holes, and has been calibrated for a number of BLR lines. The scaling relations most widely applied to {\it JWST} AGN are those by \cite{Greene05, Reines15} utilising the H$\alpha$ or H$\beta$ BLR luminosity and FWHM, typically derived from single-Gaussian fits. We refer the reader to these works for further detail and discussion of systematic uncertainties, but note here that black hole mass estimates can vary by factors of a few when different SE methods are applied.
Using such SE scaling relations, it became evident that the newly-discovered, low-luminosity early black holes fell in the supermassive category. 

However, is the SE approach appropriate to weigh distant black holes? Answering this question ideally requires independent, direct measurements of black holes masses, or a re-calibration of the SE method at high redshift. 
Currently, the most accurate high-redshift black hole masses come from interferometry resolving BLR dynamics with VLTI/GRAVITY+. So far, the masses of two quasars at $z\sim2$ and $z\sim4$ have been obtained with this method \citep{Abuter24, Davies25}. These measurements indicate that H$\alpha$-based SE methods could overestimate black hole masses by factors of $2-10$, not least due to non-virial motions within the BLR. However, these quasars are characterised by super-Eddington accretion states ($\lambda_{\rm Edd}=6-20$) which may impact the comparison.
This is supported by a recent study of five $z\sim2$ AGN comparing measurements from reverberation mapping to several SE estimators applied to {\it JWST} data \citep{Parlanti25b}. The study finds good overall agreement, in particular with an H$\beta$ calibration \cite{Vestergaard06}, except for one source accreting close to its Eddington limit for which SE methods overestimate the mass by $\sim1$~dex. 
Although limited in statistics, these findings suggest that SE methods may reliably recover masses of distant AGN in low accretion states, but may overestimate masses of highly accreting black holes \citep[see also][]{DallaBonta25, Woo26}. Conversely, if {\it JWST} AGN are accreting at super-Eddington rates, SE methods not accounting for the Eddington ratio may overestimate their black hole masses \citep[see also][]{Lupi24, Lambrides24}. It will be important to corroborate these results with larger samples for our understanding of early black hole growth. 

Specifically for LRDs, various lines of evidence suggest that SE methods systematically overestimate black hole masses. This is for instance motivated by the potential role of scattering mechanisms in the dense gas structures surrounding LRDs which could inflate the broad-line wings typically associated with the virilised BLR \citep{Chang25, Naidu25, Rusakov26}.  This can in principle be tested through observations of BLR emission in several hydrogen lines \citep[see][]{Chang25, Brazzini25}. The high covering factor of the dense, luminous gas around LRDs could also lead to a systematic overestimation of black hole masses by factors of a few when using the SE method \citep[e.g.][]{Maiolino25a}. 
For two LRDs it was shown that their bolometric luminosities are probably overestimated by 1~dex, which could impact their black hole mass estimates in the context of the gas envelope model (\citealp{Greene26}, see also \citealp{Inayoshi26}).
However, we note the case of one lensed LRD (accreting at $\lambda_{\rm Edd}\sim0.03$), for which ionised gas kinematics at few tens of parsec scales constrained a black hole mass consistent with SE estimates \citep{Juodzbalis25b}. 
We refer the reader to \cite{Inayoshi26} for further discussion of these topics.

\subsection*{The early black hole landscape}

\begin{figure}
\centering
\includegraphics[width=\textwidth]{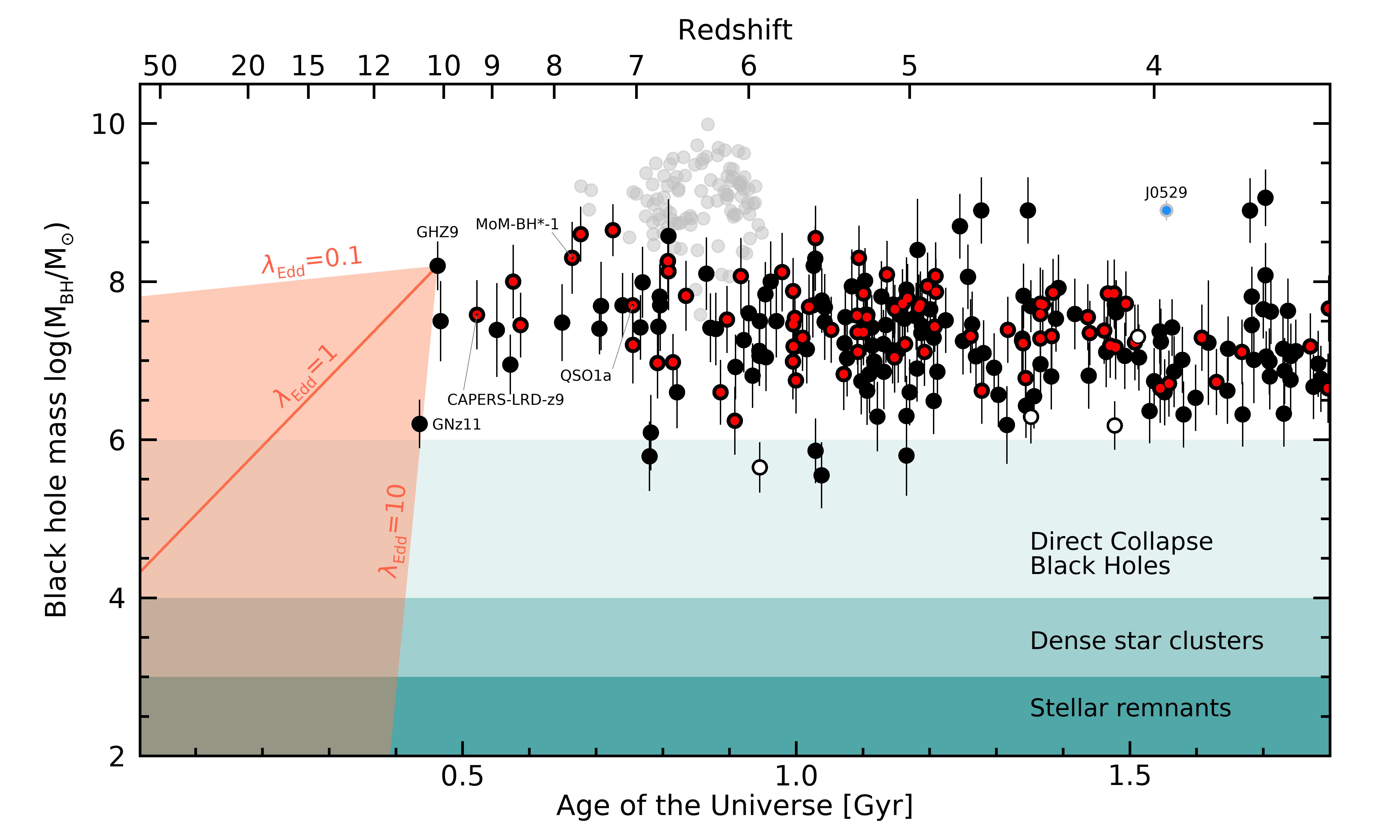}
\caption{Black hole mass as a function of age of the Universe and redshift, for the first 1.8~Gyr after the Big Bang. 
A compilation of the most distant quasars is shown in grey \citep{Fan23}. 
{\it JWST} AGN are shown in black; LRDs are marked in red. Masses are as reported in the literature, predominantly from SE methods; we note that different SE methods can give masses varying by factors of a few. To mass uncertainties below $0.3$~dex we add 0.4~dex in quadrature to account for the scatter in local scaling relations \citep[e.g.][]{Ho14}.
Four black hole candidates are marked in white \citep{Maiolino24, Uebler24c}. 
A quasar with a direct black hole mass measurement from BLR dynamics using VLTI/GRAVITY+ is indicated in blue \citep{Davies25}. 
Teal shading indicates different seed black hole regimes, while mass estimates for primordial black holes cover the full range $\lesssim10^5 M_\odot$. 
The strongest constraints on black hole seeding come from the AGN GHZ9 \citep{Napolitano25} with a mass derived from $X-$ray properties, for which we indicate growth tracks at constant Eddington ratios, with radiative efficiency $\epsilon=0.1$ and a 100~\% duty cycle. Taking the mass measurements at face value, for this and other $z\gtrsim9$ black holes an origin from light seeds growing at sub-Eddington levels can be excluded. 
We label few other sources of interest: GNz11, the most distant AGN currently known \citep{Maiolino23a}; CAPERS-LRD-z9, the most distant LRD currently known \citep{Taylor25b}; MoM-BH*-1, the LRD with the strongest Balmer break \citep{Naidu25} (non-SE mass estimates go down to $10^6 M_\odot$); QSO1a, the most distant black hole with a dynamical mass estimate \citep{Juodzbalis25b}. Other {\it JWST} AGN are compiled from the literature \citep{Kocevski23, Uebler23, Harikane23, Goulding23, Larson23, Carnall23, Uebler24b, Matthee24, Parlanti24, Lin24, Akins25, Napolitano25, Rinaldi25, Taylor25, Tripodi25b, Kiyota25, LiJ25, Juodzbalis25, Maiolino24, DEugenio25b, Uebler25, Kocevski25, Kokorev25, Greene24, deGraaff25a, Schindler25, ArevaloGonzales25, DEugenio26b, Jones26, Fei26, Lin26, ZhangJ26}.}\label{fig:mbh-z}
\end{figure}

Overall, it is clear that further data and importantly independent direct measurements are needed to robustly calibrate distant black hole masses. 
With these caveats in mind, in Fig.~\ref{fig:mbh-z} we show the masses of {\it JWST} AGN within the first 1.8~Gyr. We adopt masses directly from the literature. While the majority of data points are obtained from SE methods by \cite{Greene05, Reines15}, we note that different works differ in their spectral extraction method, fitting approaches, correction for dust attenuation, etc., which can introduce systematic differences among data sets.
A small subset of sources have black hole mass estimates based on the SE relation by \cite{Vestergaard06} or other scaling relations. The masses of few individual black holes have been determined based on their $X-$ray luminosity (\citealp{Napolitano25, Bogdan24}, but see also \citealp{Zou26}). For GNz11, the black hole mass has been estimated from SE methods using the Mg{\sc{ii}}$\lambda\lambda2796,2804$ and N{\sc{iv}}$\lambda1486$ lines, and we refer the reader to \cite{Maiolino23a} for a detailed discussion of the AGN signatures in this source \citep[see also][]{AlvarezMarquez25}.
We mark the sub-class of LRDs in red, when classified as such in the literature.
The figure highlights the new parameter space opened up through {\it JWST}, pushing towards more distant and lighter black holes.
It also visualizes an observational limitation: a scarcity of black holes below the SMBH category. Black holes with masses $\lesssim10^{5.5} M_\odot$ (based on SE methods) are likely missed in {\it JWST} observations of individual systems due to a combination of selection bias, sensitivity limitations, and spectral resolving power \citep[e.g.][]{Pacucci23, Li25, Sun25, Geris25}. Therefore, the existence of massive black holes within the first few Gyr does not exclude a dominant population of lighter black holes, including light seeds. 
However, if SE black hole masses of {\it JWST} AGN turn out to be robust within about 1~dex, then the new early black hole landscape supports the existence of moderate-to-heavy seeds and/or prolonged phases of accretion at or above the Eddington limit \citep[see e.g.][for theoretical work supporting this picture]{Schneider23}.  
For the most distant black holes discovered so far, an origin from light seeds and continuous sub-Eddington growth can be ruled out.

\section*{Host galaxies and surroundings}

Considering the high black hole masses inferred for {\it JWST} AGN, we next turn our attention to the scalings with galaxy properties. Turns out, not only are {\it JWST} black holes supermassive, they also appear overmassive compared to their galaxies' stellar mass, when considering local scaling relations \citep[e.g.][]{Uebler23, Harikane23, Maiolino24, Juodzbalis25, Gupta26}. Are stellar masses robust? 
Stellar masses are typically derived from SED fitting of the broad-band photometry and/or integrated spectra. These estimates come with considerable uncertainties because the AGN contribution to the overall SED is often difficult to determine. To mitigate this, some works derive stellar masses based on spatially-resolved SED fitting, or after a photometric decomposition of an AGN and galaxy component \citep[e.g.][]{Ding23, Harikane23, Juodzbalis25}. Works comparing stellar mass estimates for AGN based on different setups, codes or decompositions show those can differ by more than 0.75~dex, while statistical uncertainties for individual methods are typically in the range $0.3-0.4$~dex \citep[e.g.][]{Juodzbalis25, Gupta26}. For a detailed discussion of the assumptions and systematics involved, we refer the reader to the above works, but we here caution that stellar masses can be similarly uncertain as black hole masses, and in particular for LRDs \citep{Gupta26}. Important in the context of black hole-galaxy scaling relations though, we note that there is no indication that stellar masses are systematically underestimated by 1~dex.

\begin{figure}
\centering
\includegraphics[width=0.49\textwidth]{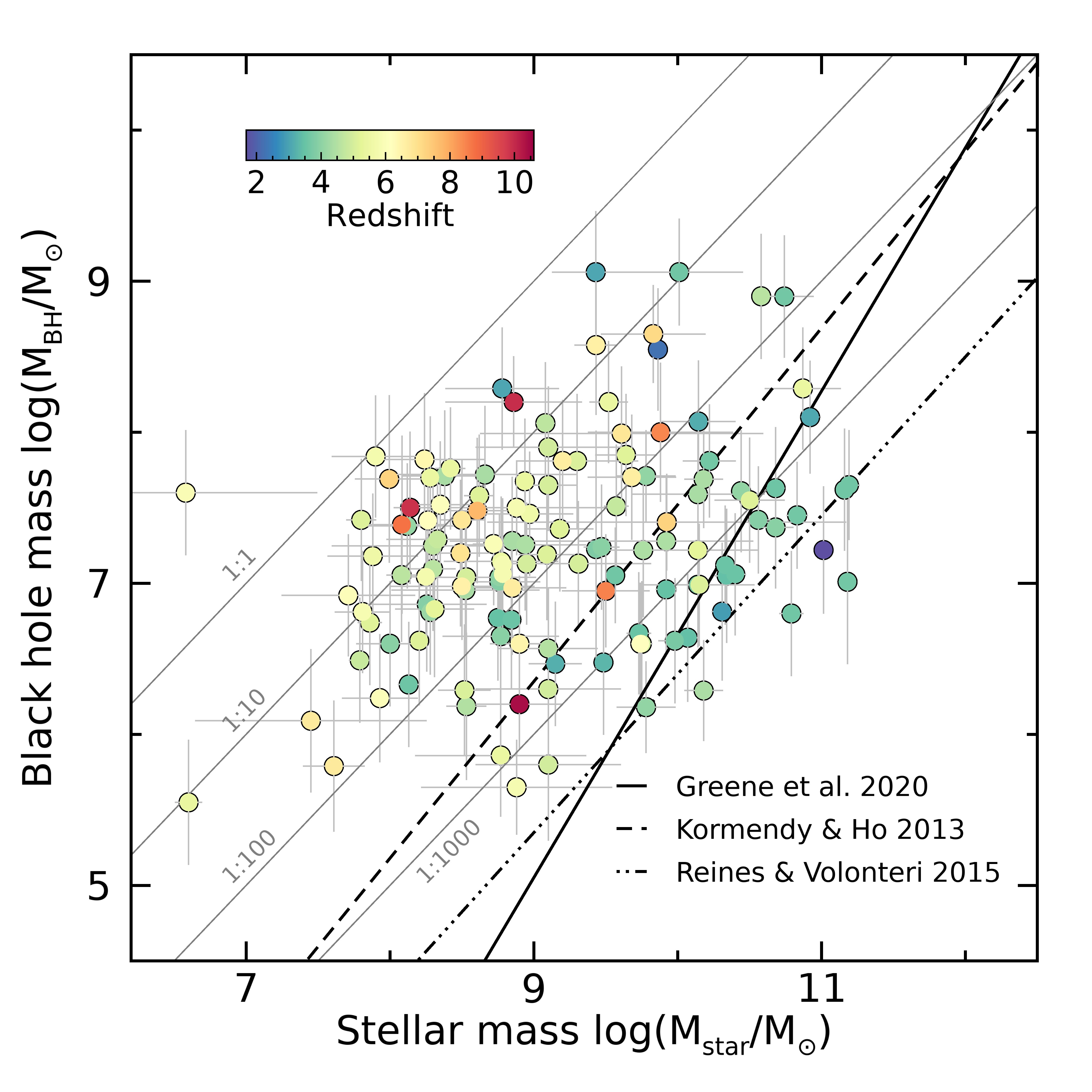}
\includegraphics[width=0.49\textwidth]{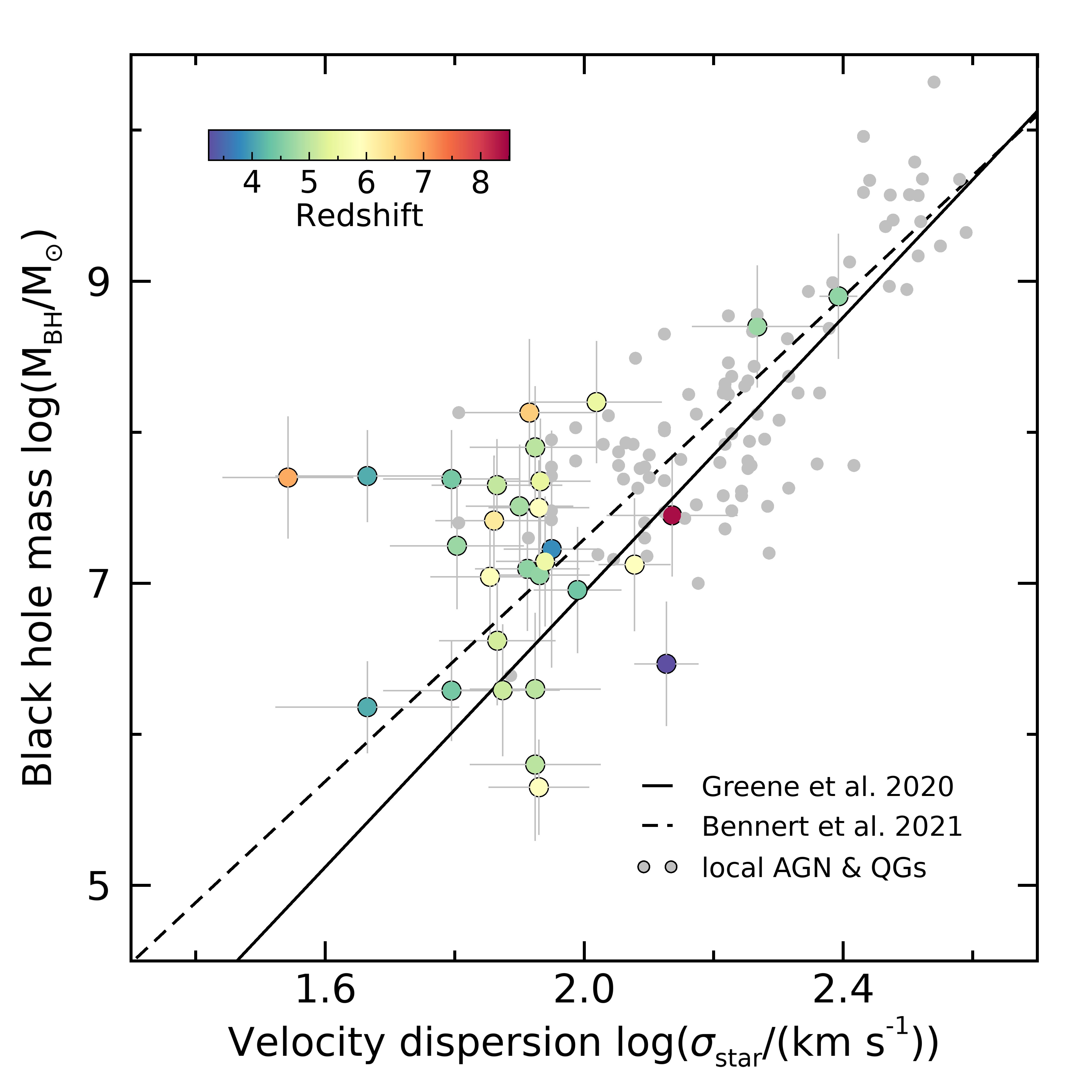}
\caption{Black hole mass as a function of galaxy properties. {\it Left:} black hole mass as a function of galaxy stellar mass. {\it Right:} black hole mass as a function of integrated stellar velocity dispersion. Masses are as reported in the literature. For sources with integrated gas dispersion measurements, we convert those to integrated stellar velocity dispersion using the empirical scaling of \cite{Bezanson18b}.
At fixed stellar mass, {\it JWST} AGN have black hole mass estimates (from SE methods) about two orders of magnitude higher compared to local scaling relations \citep{Kormendy13, Reines15, Greene20}. In the black holes mass -- stellar velocity dispersion plane, {\it JWST} AGN scatter around the lower end of local scaling relations for AGN and quiescent galaxies (`QGs') \citep{Kormendy13, Bennert21}.
{\it JWST} AGN are compiled from the literature: \citep{Uebler23, Maiolino23a, ZhangJ26, Rinaldi25, Goulding23, Napolitano25, Akins25, Larson23, Tripodi25b, Kiyota25, Juodzbalis25, Maiolino24, DEugenio26, Uebler25, Kocevski25, Wang25, Harikane23, Carnall23, Gupta26, Schindler25, ArevaloGonzales25, Fei26, LiJ25} (left) 
and \citep{Uebler23, Uebler25, Parlanti24, Maiolino24, Juodzbalis25, DEugenio26, Juodzbalis25b, DEugenio25a, Jones26, DEugenio26b, Carnall23} (right).
}\label{fig:mbh-mstar-sig}
\end{figure}

In the left panel of Fig.~\ref{fig:mbh-mstar-sig}, we show black hole mass as a function of stellar mass for {\it JWST} AGN, colour-coded by redshift. At fixed stellar mass, the majority of {\it JWST} AGN are offset towards higher black holes masses. For the bulk of the population, SE methods would have to overestimate black hole masses by about two orders of magnitude to bring them into agreement with local scaling relations. The figure also shows tentative evidence for a possible redshift trend, such that more distant AGN show a larger offset from local relations, as noted by \cite{Juodzbalis25, JonesB25, Gupta26}.
As mentioned before, the existence of black holes more massive than expected based on local scaling relations, which appears robust even after considering the uncertainties in mass measurements discussed above, does not imply that all early black holes are overmassive. Lower-mass black holes are likely missed by {\it JWST}, and even the overmassive black holes could remain in better agreement with local scalings if scatter in the relations increases substantially with redshift \citep{Li25, Silverman25, Sun25, Ziparo26, Gupta26}.
We further note here that the general finding of overmassive black holes is not new, but was already reported for several distant quasars, some low-luminosity AGN at intermediate redshifts, and recently also for some inactive black holes in the local Universe, in relation to various galaxy properties \citep[e.g.][]{WangR10, Mezcua23, deNicola25}.

While many {\it JWST} black holes appear overmassive when related to stellar mass, this is not the case for some other galaxy properties commonly related to black hole mass. Specifically, measures of the integrated velocity dispersion combined with black hole mass follow local relations \citep{Maiolino24, Juodzbalis25, Uebler25}. 
Because the integrated velocity dispersion is a measure of the total potential, this could indicate that in terms of total mass distant {\it JWST} AGN are not particularly overmassive. -- We note that direct measurements of dynamical mass from forward-modelling of ionised gas kinematics are challenging due to the compactness of {\it JWST} AGN. Most works use estimates from unresolved kinematics to approximate $M_{\rm dyn}$, generally finding better agreement with local relations compared to $M_{\rm star}$ \citep[e.g.][]{Juodzbalis25}. --
However, even for the relation with velocity dispersion statistics are still limited due to the requirement of high spectral resolution data to measure the velocity dispersion. Another caveat when comparing to local relations is the conversion of ionised gas to stellar velocity dispersion, which is studied only out to $z\sim1$ \citep{Bezanson18b}.
In the right panel of Fig.~\ref{fig:mbh-mstar-sig}, we show black hole mass as a function of galaxy stellar velocity dispersion. {\it JWST} AGN are in broad agreement with local scaling relations.

Other galaxy properties such as metallicities or sizes are not as systematically studied.
While some AGN are found embedded in extended, regularly rotating galaxies \citep[e.g.][]{Parlanti24, Uebler24c}, most are found in compact systems with effective radii of a few hundred parsec or less \citep[e.g.][]{Hainline25}. 
At first glance, this appears different to quasars and $X-$ray-selected AGN, whose rest-frame optical emission generally reveals complex, extended structures, neighbours and mergers \citep[e.g.][]{Wylezalek22, Perna23, Marshall23, Decarli24, Zamora24, Perna25a, Marshall25a, Wolf26}. 
However, faint, blue emission or companions are visible around many lower-luminosity {\it JWST} AGN \citep[e.g.][]{Uebler23, Uebler24b, Matthee24, Chen25b, Baggen26, Barger26}. 
So far, few {\it JWST} AGN are found in more complex environments containing several, partly more massive, neighbours \citep[][]{DEugenio26, Tozzi26, Iskikawa26}.

A number of high-redshift systems have been identified with NIRSpec-IFS containing multiple massive black holes \citep[][]{Perna23, Uebler24b, Zamora24, Uebler25, Perna25a, Perna26}.
At few kpc to sub-kpc separations, these black holes are expected to merge within a Hubble time. Such systems can serve as benchmarks for theoretical models of galaxy and black hole growth in the early Universe \citep[e.g.][]{DeRosa19}, and provide forecasts for the next-generation gravitational waves observatory {\it LISA} which will be sensitive to massive black hole mergers across cosmic time. 
The key stages preceding the coalescence of black holes through the emission of gravitational waves -- galaxy mergers, dynamical friction, binary hardening -- are critically influenced by galaxy environments and physical properties, such as gas fraction, dynamics, stellar density, or galactic substructures \citep[see][for a detailed review]{AmaroSeoane23}.
Increasing statistics on the environments and properties of multiple black holes systems in the early Universe is a key observational goal for the coming years.

Several programs have attempted to resolve the host galaxies of compact {\it JWST} AGN and LRDs at redder wavelengths ($\lambda_{\rm rest}>1\mu$m) using MIRI \citep[e.g.][]{Lyu24}, ALMA and NOEMA. Interestingly, the typically bright [C{\sc{ii}}]$\lambda158\mu$m line remains largely undetected \citep[e.g.][]{Akins25, Setton25, Xiao25, Mazzolari26b}, which could be due the ionizing radiation of the AGN, to high densities or low metallicities. For the few sources with reported (tentative) detections the emission could be associated with neighbours \citep[see][]{Mazzolari26b}, and higher-resolution follow-up is required to test this.
Specifically for LRDs, current data in the rest-frame mid-to-far-infrared place limits on dust masses ($\lesssim10^6 M_\odot$ for $\sim100$~K dust at $z\sim6$) with little contribution from hot dust \citep[see][for further details]{Williams24, Setton25, Casey25, Delvecchio25, Leung25}.

Together with the compactness of LRDs, these findings suggest that LRDs could be without a host galaxy \citep[e.g.][]{Hviding25, Chen25a}. If true, this would be extremely important, as it would imply that LRDs could be directly related to the formation of black holes. 
Some works, mainly from a theoretical perspective, suggested that LRDs could be DCBHs \citep{Cenci25, Baggen26, Pacucci26}. It was speculated that the Lyman-Werner radiation required to form DCBHs could have been supplied by low-mass neighbours \citep{Baggen26}, in line with the synchronized pair scenario \citep[e.g.][]{Visbal14}. Others argued that the mildly attenuated radiation of DCBHs can explain properties of LRDs such as the weakness of metal lines and the spectral shape, requiring virtually no contribution from stars \citep{Pacucci26}. It has also been suggested that LRDs could be the descendants of primordial black holes \citep[e.g.][]{Maiolino26a, Zhang26}, similarly supporting a heavy-seed formation channel.
From the observational perspective, there is one intriguing LRD with unresolved continuum emission despite being magnified by a factor $\mu\approx6.2$: QSO1a at $z\sim7$ \citep{Furtak24}. NIRSpec-IFS observations revealed a dominant black hole, at least twice as massive as the galaxy \citep{Juodzbalis25b}, which in turn is characterised by very low metallicity ($Z\sim4\times10^{-3}~ Z_\odot$) \citep{Maiolino26a}. These observed properties are best explained through a heavy seed formation channel \citep{Maiolino26a, Juodzbalis25b}.

\section*{Outlook}\label{}

{\bf Inactive black holes. } 
This article is concerned with distant, active black holes. Many SMBHs in the local Universe are inactive and have been detected through their gravitational impact on surrounding stars. At high redshift, such measurements are largely out of reach with current facilities. It is thus noteworthy that a mass measurement of a six billion solar mass black hole in a quiescent, lensed galaxy at $z=1.95$ was achieved through stellar dynamics using {\it JWST} data \cite{Newman26}. 
The superior spatial and spectral resolution of upcoming instruments such as ELT/MICADO will enable mass measurements for statistical samples of inactive black holes at cosmological distances.

{\bf The future. }  
As discussed in this article, current mass estimates of distant black holes are subject to significant uncertainties. 
Important progress will be made through more extensive cross-calibrations of SE methods and direct mass measurements. This can be achieved today with VLTI/GRAVITY+, and in the near future with instruments like ELT/MICADO. Higher spatial resolution observations will enable a better characterisation of host galaxy properties, and higher spectral resolution helps to decompose the BLRs of lighter black holes from interstellar medium or outflow emission. 
In addition, several ongoing {\it JWST} programs will improve constraints on black hole masses by tracing the BLR emission across different electron transitions and different atoms.
Across the electromagnetic spectrum, upcoming facilities promise to widen our understanding of distant AGN. For instance, thanks to its area--sensitivity combination the planned $X-$ray mission {\it NewAthena} will be able to detect statistical samples of (un)obscured AGN up to $z\sim6-7$ \citep[e.g.][]{Cruise25}.
At radio wavelengths, the Square Kilometer Array Observatory will enable the detection of thousands of AGN within the first billion years \citep[e.g.][]{Mazzolari24a}. 
With their increased sensitivities compared to previous facilities, these observatories will sharpen our view on {\it JWST} AGN which are currently characterised by $X-$ray and radio upper limits.
Pulsar Timing Arrays are starting to constrain the high-mass end of the SMBH mass function up to $z\sim1.5$ \citep{Agazie23, EPTA24}. In a decade from now, {\it LISA} will provide us with a black hole mass function in the range $\sim10^{4-7} M_\odot$ out to $z>20$ through the signatures of merging black holes.

\backmatter

\bmhead{Acknowledgements}

I am grateful to Dale Kocevski for help in reproducing Fig.~1, and to Giovanni Mazzolari, Steve Finkelstein, Roberto Maiolino, Taro Shimizu, and Frank Eisenhauer for valuable comments and discussions.
I acknowledge support by the Max Planck Society through the Lise Meitner Excellence Program, and funding by the European Union (ERC APEX, 101164796). Views and opinions expressed are however those of the author only and do not necessarily reflect those of the European Union or the European Research Council Executive Agency. Neither the European Union nor the granting authority can be held responsible for them.

\bibliography{sn-bibliography}

\section*{Statement of Competing Interests}
The author declares no competing interests.

\end{document}